\documentclass[11pt]{article}
\usepackage{graphicx,color}

\usepackage{epstopdf}
\usepackage{tikz}
\usepackage{amsmath,amssymb,amscd,latexsym,amsthm,mathrsfs}

\newtheorem{theo}{Theorem}

\ifx\pdfoutput\undefined\else

\def\@begintheorem#1#2{\par\bgroup{\sc #1\ #2. }\it\ignorespaces}
\def\@opargbegintheorem#1#2#3{\par\bgroup{\sc #1\ #2\ (#3). }\it\ignorespaces}
\def\@endtheorem{\egroup\par}

\def\Pr{{\it Proof:  }}
\renewcommand{\author}[1]{\large\rm #1\\ \bigskip}
\newcommand{\address}[1]{{\normalsize\it #1\\}\bigskip}
\renewcommand{\title}[1]{\bigskip\bigskip\Large\bf #1\bigskip\bigskip\\}

\begin{document}
\vglue .3cm
\begin{center}
\title{Pulling adsorbed self-avoiding walks from a surface}
\author{Anthony J. Guttmann\footnote[1]{email:
                {\tt tonyg@ms.unimelb.edu.au}}, I. Jensen\footnote[2]{email:
                {\tt iwan@ms.unimelb.edu.au}}}
\address { ARC Centre of Excellence for\\
Mathematics and Statistics of Complex Systems,\\
Department of Mathematics and Statistics,\\
The University of Melbourne, Victoria 3010, Australia}
 \author{ and S. G. Whittington\footnote[3]{email:
                {\tt swhittin@chem.utoronto.ca}}}
\address{Department of Chemistry, University of Toronto, Toronto, Canada}
\end{center}

\setcounter{footnote}{0}


\begin{abstract}
We consider a self-avoiding walk model of polymer adsorption where the
adsorbed polymer can be desorbed by the application of a force, concentrating on the
case of the square lattice. Using series analysis methods we investigate the behaviour 
of the free energy of the system when there is an attractive potential $\epsilon$ with the 
surface and a force $f$ applied at the last vertex, normal to the surface, and extract the 
phase boundary between the ballistic and adsorbed phases. We believe this to be exact 
to graphical accuracy. We give precise estimates of the location of the transition from the free
phase to the ballistic phase, which we find to be at $y_c=\exp(f/k_B T_c)=1,$ 
and from the free phase to the adsorbed phase, which we estimate to be at 
$a_c=\exp(-\epsilon/k_B T_c)=1.775615 \pm 0.000005$. In addition we
prove that the phase transition from the ballistic to the adsorbed phase is first order.
 
\end{abstract}


\section{Introduction}
\setcounter{equation}{0}
\label{sec:intro}

The theory of polymer adsorption \cite{DeBell,Rensburg2000} has a long history \cite{Rubin,Silberberg1962}.
For linear polymers a variety of models have been considered including random walks
\cite{Hammersley1982,Rubin,Binder2012}, directed and partially directed walks
\cite{Forgacs,Rensburg2003,Whittington1998} and self-avoiding walks 
\cite{Batchelor1995,Beaton2012,Guim1989,HTW1982,Hegger1994,Rensburg1998,Rensburg2004}.  In this paper we shall 
be concerned with the self-avoiding walk model for which there are a few rigorous results
\cite{HTW1982,Rensburg1998} as well as extensive numerical investigations (see for 
instance \cite{Beaton2012,Guim1989,Hegger1994,Rensburg2004}).

The invention of micro-manipulation techniques such as atomic force 
microscopy (AFM) \cite{Zhang2003} which allow adsorbed polymer molecules to be pulled off a surface
\cite{Haupt1999} has led to the development of theories of adsorbed polymers subject
to a  force
\cite{Krawczyk2005,Orlandini1999,Owczarek2010,Skvortsov2009,Binder2012}.  
Much of this work has focussed on random,
directed and partially directed walk models but there has been some numerical work on the 
self-avoiding walk model \cite{Iliev2013,Krawczyk2005,Mishra2005} and a recent rigorous treatment
\cite{JvRW2013} which establishes the existence of a phase boundary between an adsorbed 
phase and a ballistic phase when the force is applied normal to the surface.  In this paper we use exact enumeration and series analysis 
techniques to identify this phase boundary for self-avoiding walks on the square
lattice.  We also make precise estimates of the critical points for adsorption with no force and 
for the transition to the ballistic phase with no surface interaction, and various 
relevant critical exponents.  For a brief discussion of critical exponents appearing in this problem
see \cite{Batchelor1995} or \cite{Guim1989}.


\section{Definitions and review of rigorous results}
\setcounter{equation}{0}
\label{sec:defns}

Consider the square lattice ${\mathbb Z}^2$ where the vertices have 
integer coordinates.  We write $(x_i,y_i)$, 
$i=0,1,2, \ldots n$ for the coordinates of the $i$-th vertex of an 
$n$-step self-avoiding walk on ${\mathbb Z}^2$. 
The number of $n$-step self-avoiding walks  from the origin is denoted by
$c_n$. It is known that $\lim_{n\to\infty} n^{-1}  \log c_n 
= \log \mu$ exists \cite{HM54}, where $\mu$ is the 
\emph{growth constant} of self-avoiding walks on this lattice.

A \emph{positive walk} is a self-avoiding walk on ${\mathbb Z}^2$ that starts 
at the origin and is constrained to have $y_i \ge 0$ for all 
$0 \le i \le n$.  The number of $n$-step positive walks from 
the origin is denoted by $c_n^+$.  It is known that 
$\lim_{n\to\infty} n^{-1} \log c_n^+ = \log \mu$ \cite{Whittington1975}.
Vertices of a positive walk with $y_i=0$ are \emph{visits} to the surface although, 
by convention, the vertex at the origin is not 
counted as a visit.  We say that the walk has \emph{height} $h$ if for the last vertex  
 $y_n=h$. The number of positive walks of $n$-steps from the 
origin, with $v$ visits and height $h$ is denoted by $c_n^+(v,h)$.  The corresponding partition
function is 
\begin{equation}
C_n(a,y) = \sum_{v,h} c_n^+(v,h) a^v y^h.
\end{equation}
If $\epsilon$ is the energy associated with a visit and $f$ is the force applied
at the last vertex, normal to the surface,
\begin{equation}
a= \exp[-\epsilon /k_B T] \quad \mbox{and} \quad y=\exp[f/k_B T]
\label{eqn:physvar}
\end{equation}
where $k_B$ is Boltzmann's constant and $T$ is the absolute temperature.  If no force is 
applied $y=1$ and the appropriate partition function is $C_n(a,1)$ while if there
is no interaction with the surface $a=1$ and the appropriate partition function
is $C_n(1,y)$.  

It is known \cite{HTW1982} that the limit
\begin{equation}
\lim_{n\to \infty} n^{-1} \log C_n(a,1) \equiv \kappa (a)
\label{eqn:kappa}
\end{equation}
exists and that $\kappa (a) $ is a convex function of $\log a$.  There exists a value 
of $a = a_c^o > 1$ such that $\kappa (a) = \log \mu $ for $a \le a_c^o$ and $\kappa(a)$
is strictly monotone increasing for $a > a_c^o$.  
Therefore the free energy $\kappa (a)$ is non-analytic
at $a=a_c^o$ \cite{HTW1982} and this corresponds to the adsorption transition in the absence
of a force.  For $a > a_c^o$, in the adsorbed phase, 
\begin{equation}
\lim_{n\to\infty} \frac{\langle v \rangle}{n} > 0
\end{equation}
while, for $a<a_c^o$,  $\langle v \rangle = o(n)$.  Here $\langle \cdots \rangle$ denotes expectation.

Similarly it is known \cite{Rensburg2009} that the limit
\begin{equation}
\lim_{n\to \infty} n^{-1} \log C_n(1,y) \equiv \lambda (y)
\label{eqn:lambda}
\end{equation}
exists and $\lambda (y) $ is a convex function of $\log y$.   There is a critical point
$y_c^o \ge 1$ such that  $\lambda (y) = \log \mu$ for $y \le y_c^o$ and $\lambda (y) $ is 
strictly monotone increasing for $y > y_c^o$ \cite{Rensburg2009}.  The critical point corresponds
to a transition from a free phase where $\langle h \rangle = o(n)$ to a ballistic phase where 
\begin{equation}
\lim_{n\to\infty} \frac{ \langle h \rangle }{n} > 0.
\end{equation}
There are good reasons to believe \cite{IoffeVelenik,Rensburg2009} that $y_c^o=1$.

For the full two variable model it has recently been shown \cite{JvRW2013} that the limiting free energy
\begin{equation}
\psi (a,y) = \lim_{n\to\infty} n^{-1} \log C_n(a,y)
\end{equation}
exists.  $\psi(a,y)$ is a convex function of $\log a$ and $\log y$ (\emph{i.e.} convex as a surface) and 
\begin{equation}
\psi(a,y) = \max[\kappa (a), \lambda (y)].
\label{eqn:psi}
\end{equation}
This implies that there is a \emph{free phase } when $a < a_c^o$ and $y< y_c^o$ 
where $\langle v \rangle = o(n)$ and $\langle h \rangle = o(n)$ and a strictly monotone curve $y=y_c(a)$ through the point $(a_c^o,y_c^o)$ separating two  phases:
\begin{enumerate}
\item
an \emph{adsorbed phase} when $a > a_c^o$ and $y < y_c(a)$, and
\item
a \emph{ballistic phase} when $y >\max[y_c^o, y_c(a)]$.
\end{enumerate}
Moreover, for the square lattice, $y_c(a)$ is asymptotic to $y=a$ as $a \to \infty$.

\section{Exact enumerations \label{sec:enum}}

The algorithm we use to enumerate self-avoiding walks (SAW) on the square lattice builds on the 
pioneering work of Enting \cite{Enting80} who enumerated square lattice 
self-avoiding polygons (SAP) using the finite lattice method. More specifically 
our algorithm is based in large part on the one devised by Conway, Enting and 
Guttmann \cite{Conway93a} for the enumeration of SAWs. Many details of our
algorithm can be found in \cite{Jensen04}.  
All of the above transfer matrix (TM) algorithms are based on keeping track of the way 
partially constructed SAW are connected to the left of a cut-line bisecting
the given finite lattice (rectangles in the case of the square lattice). 
Recently Clisby and Jensen \cite{Clisby12} devised
a new and more efficient implementation of the  transfer-matrix
algorithm for self-avoiding polygons.   In that implementation we took a new approach and instead 
kept  track of how a partially constructed SAP must connect up to the right of the
cut-line. Jensen  extended this approach to the enumeration of SAW \cite{Jensen13}.
Here we briefly describe how this algorithm can be amended to enumerate SAW 
configurations for the problem we study in this paper.
 
The first terms in the series for the SAW generating 
function can be calculated using transfer matrix techniques to count 
the number of walks in rectangles $W$ unit cells wide and $L$ cells long. 
Any walk spanning such a rectangle 
has a  length of at least $W+L$ steps. By adding the contributions 
from all rectangles of width $W \leq W_{\rm max}$  (where the choice of 
$W_{\rm max}$ depends on available computational resources) and length 
$W \leq L \leq 2W_{\rm max}-W+1$  the number of walks per vertex of an 
infinite lattice is obtained correctly up to length $N=2W_{\rm max}+1$.

\begin{figure}[ht]
\begin{center}
\includegraphics[scale=0.7]{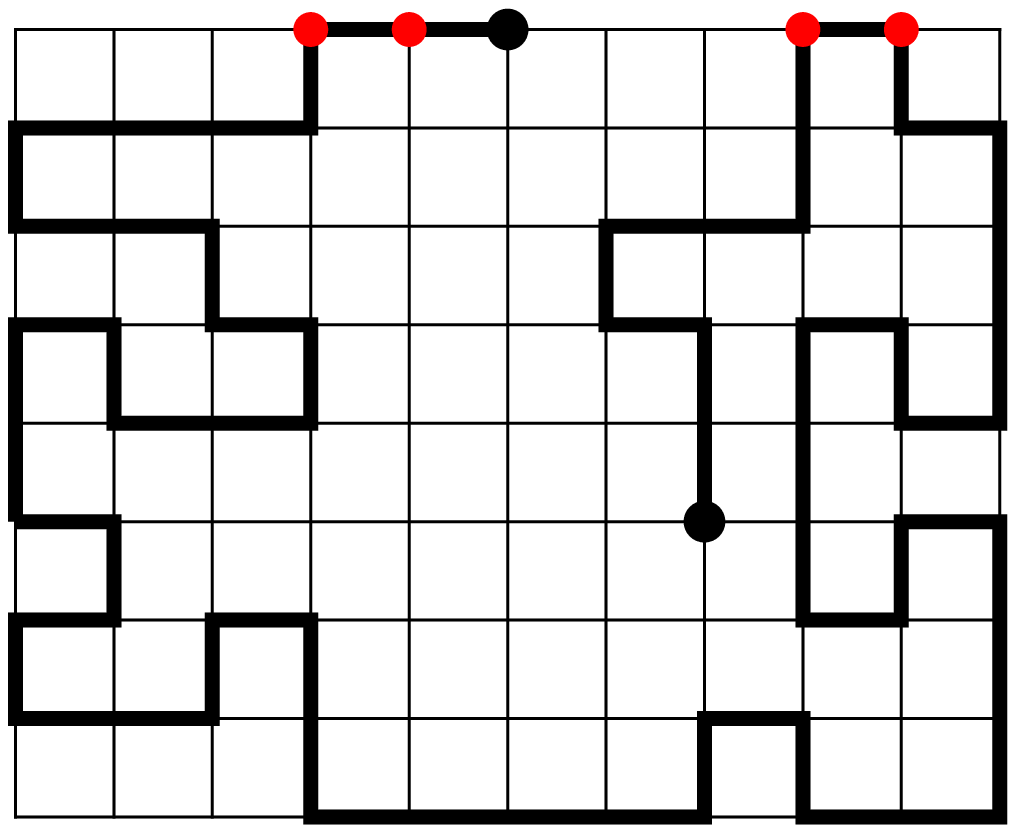}
\end{center}
\caption{\label{fig:sawex}
 An example of a   self-avoiding walk on a $10\times 8$ rectangle. The walk is tethered to the surface,
 has  the end-point at $h=5$ and four vertices (other than the start-point)  in the surface.}
\end{figure}

\begin{figure}[ht]
\begin{center}
\includegraphics[scale=0.9]{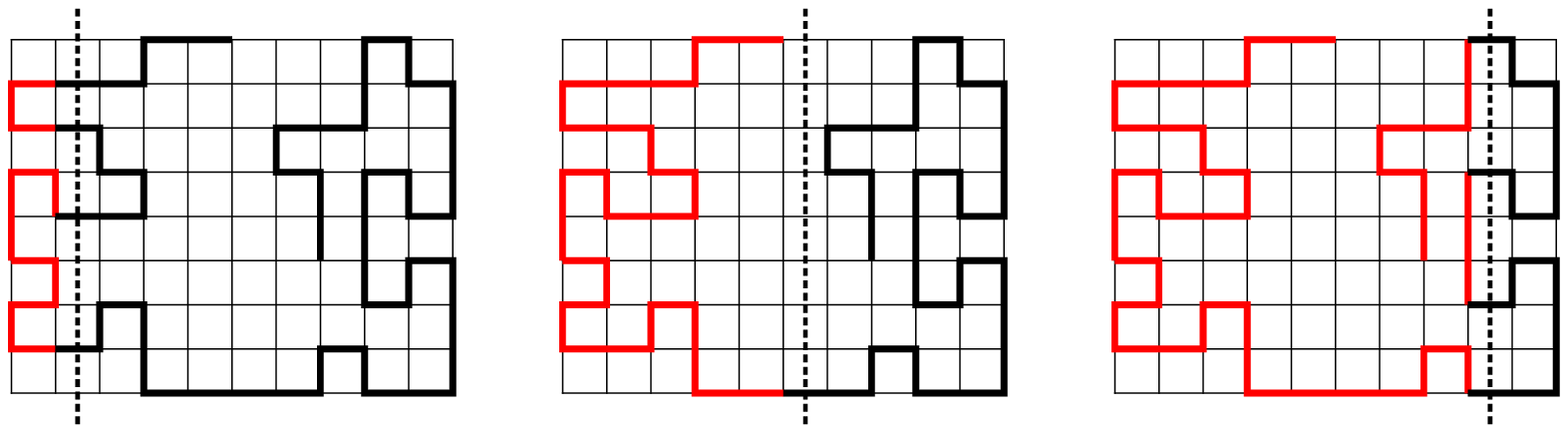}
\end{center}
\caption{\label{fig:sawcut}
 Examples of cut-lines through the SAW of Fig.~\ref{fig:sawex} such that
 the signature of  the yet to be completed section to the right of the cut-line 
 (black lines) contains, respectively, two, one and no free edges.}
\end{figure}

The basic idea of the  algorithm can best be illustrated by
considering the specific example of a SAW given in figure~\ref{fig:sawex}.
Clearly any SAW is topologically equivalent to a line and therefore
has exactly two end-points. If we cut the SAW by a vertical line as shown 
in figure~\ref{fig:sawcut}  (the dashed line) we see that the SAW is broken into 
several pieces to the left and right of the cut-line. On {\em either} side of the cut-line 
we have a set of  arcs connecting two edges on the cut-line and at most  
two line pieces connected to the end-points of the SAW.
As we move the cut-line from left to right we  prescribe what must happen 
in the future, that is how edges are to be connected to the right of the cut-line so as to
form a valid SAW.  Each end of an arc is assigned one of two labels 
depending on whether it is the lower or upper end of an arc. Any configuration along the
cut-line can thus be represented by a set of edge states
$\{\sigma_i\}$, where

\begin{equation}\label{eq:states}
\sigma_i = \left\{ \begin{array}{rl}
0 &\;\;\; \mbox{empty edge}, \\ 
1 &\;\;\; \mbox{lower edge}, \\
2 &\;\;\; \mbox{upper edge}. \\
3 &\;\;\; \mbox{free edge}. \\
\end{array} \right.
\end{equation}
\noindent
If we read from the bottom to the top, the configuration or signature $S$ along the 
cut-lines of the SAW in figure~\ref{fig:sawcut} are, respectively, $S=\{030010230 \}$, 
 $S=\{300000000\}$, and $S=\{102001002 \}$.
Since crossings are not permitted this encoding uniquely describes 
how the occupied edges  are connected. 

The most efficient implementation of the TM algorithm generally involves moving 
the cut-line in such a way as to build up the lattice vertex by vertex. 
The sum over all contributing graphs is calculated as the cut-line is moved through the lattice. 
For each configuration of occupied or empty edges  along the intersection we maintain a 
generating function $G_S$ for partial walks with signature $S$. In exact enumeration studies 
such as this $G_S$ is a truncated  polynomial  $G_S(x,a)$ where $x$ is conjugate to the
number of steps and $a$ to the number of visited vertices in the surface.  
In a TM update each source  signature $S$ (before the boundary is moved) gives rise 
to a  few new target signatures  $S'$ (after the move of the boundary line) 
as $k=0, 1$ or 2 new edges are inserted with $m=0$ or 1 surface visits
leading to the update  $G_{S'}(x,a)=G_{S'}(x,a)+x^ka^mG_S(x,a)$. Once a signature $S$ 
has been processed it can be discarded.

 Some minor changes to the basic algorithm described in 
\cite{Jensen13} are required in order to enumerate the SAW configurations for the problem 
we study in this paper. Since we are moving the cut-line so as to add one vertex at a time
we have complete control over the placement of the end-points of the SAW.
In particular, grafting the SAW to the surface can be achieved by forcing the SAW to have 
a free end (the start-point)  on the top of the rectangle. In enumerations of unrestricted SAW one 
can use symmetry to restrict the TM calculations to rectangles with  $W\leq N/2+1$ and $L\geq W$ by 
counting contributions for rectangles  with $L>W$ twice. The grafting of the start-point to the wall 
breaks the symmetry and we have to consider all rectangles with $W\leq N+1$. The number of 
signatures one must consider grows exponentially with $W$. Hence we must minimize the length of 
the cut-line to obtain an optimal algorithm. To achieve this the TM calculation on the set of rectangles is broken 
into two sub-sets with $L\geq W$ and $L<W$, respectively. The calculations for the 
sub-set with $L\geq W$ is done as outlined above. In the calculations for the  
sub-set with $L<W$ the boundary line is chosen to be horizontal (rather than vertical) so 
it cuts across at most  $L+1$ edges. Alternatively, one may view the calculation for the second sub-set 
as a TM algorithm for SAW with start-point on the left-most border of the rectangle. 
To keep track of the height $h$ of the end-point we simply specify that it must be placed in a 
row (or column) $h$ lattice-units from the surface and we then repeat the calculation for all the  possible values
of $h$.

We calculated the number of SAW up to length $n=59$. The calculation was
performed in parallel using up to 16 processors, a maximum of  some 40GB of memory 
and using a total of just under 6000 CPU hours (see \cite{Jensen04} 
for details of the parallel algorithm).

\section{Results}
\label{sec:results}

In this Section we describe the results from series analysis, chiefly using
differential approximants \cite{GJ09}.  We first discuss the $y$-dependence of the free energy
$\lambda (y)$ when there is no surface interaction, then the $a$-dependence of the 
free energy $\kappa (a)$ when there  is no applied force and finally the two variable 
free energy $\psi (a,y)$ when there is both a surface interaction and a force.

\subsection{No surface interaction. $a=1.$}\label{sec:lambda}

\begin{figure}
\centering
\includegraphics[scale =0.5] {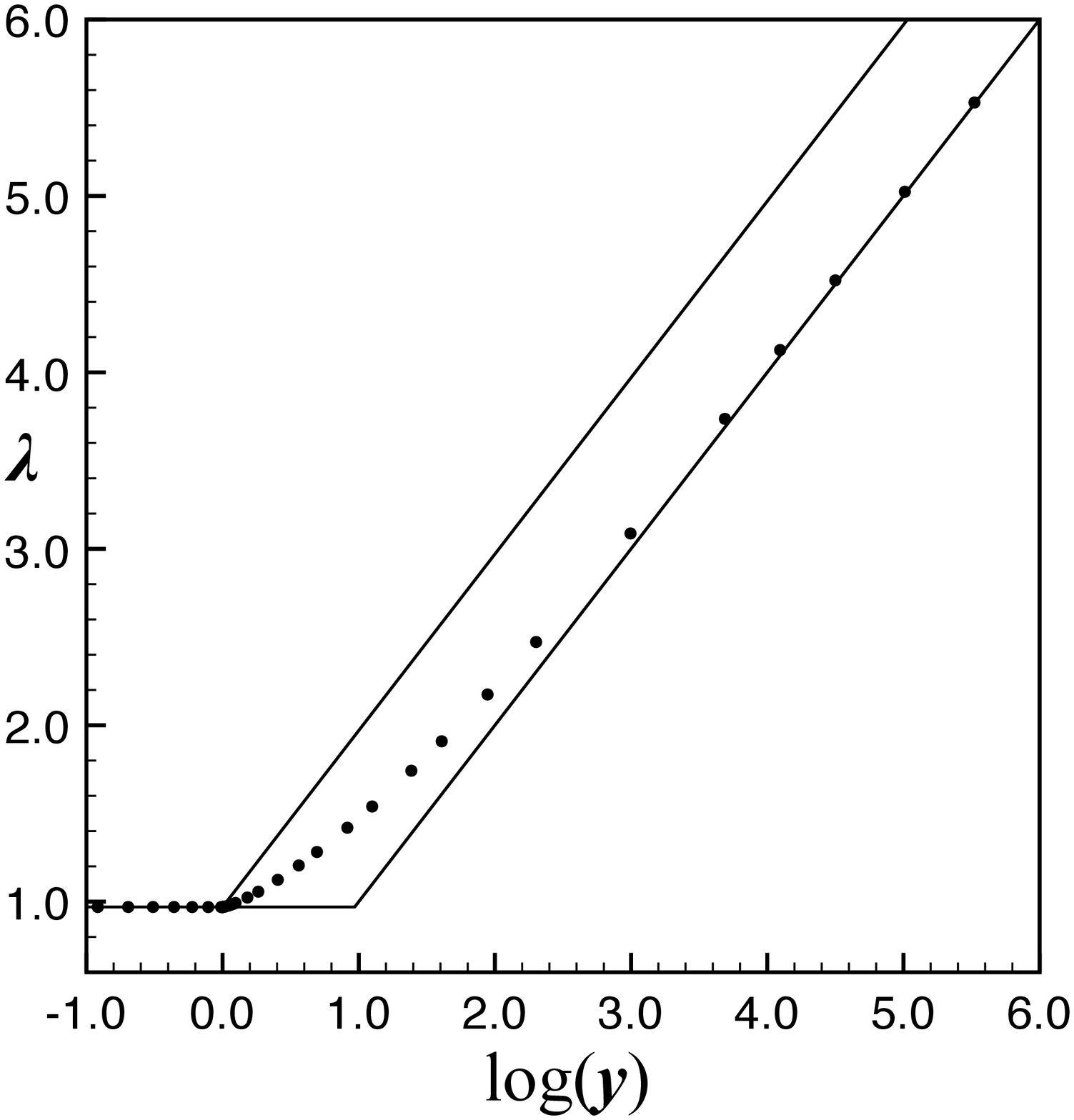}
 \caption{The $y$-dependence of the free energy $\lambda (y)$. The straight lines 
 indicate the exact lower and upper bounds, 
 where for $\log y < 0$ we know that $\lambda (y) = \log \mu$ while for $ \log y > 0$ we know that
$ \max[\log \mu, \log y] \le \lambda (y) \le \log \mu + \log y$.}
 \label{fig:lambda}
\end{figure}

If we write 
\begin{equation}
H(x,y) = \sum_n C_n(1,y) x^n=\sum_n e^{\lambda(y) n + o(n)} x^n
\end{equation}
then $H(x,y)$ will be singular at $x=x_c(y) = \exp[-\lambda (y)]$
and, close to this singularity,  $H(x,y)$ is expected to behave as 
\begin{equation}
H(x,y) \sim \frac{A}{[x_c(y)-x]^{\gamma (y)}}
\end{equation}
where $\gamma(y)$ is a critical exponent whose value depends on the value of $y$.

In  table~\ref{tab:a1}  below we give the results of an analysis of the series  $H(x,y)$ for various values of $y$.
The resulting estimates of the free energy $\lambda (y) = -\log x_c$ are plotted in figure~\ref{fig:lambda}.
The series were analysed using second and third order differential approximants \cite{GJ09}. At $y=1$ the series is well behaved and has critical point $1/\mu$ with exponent $\gamma_1=61/64,$ the exponent for terminally-attached self-avoiding walks (TASAW), as one expects \cite{Ca84}. 
For $y$ just below 1 the series are quite difficult to analyse. Estimates of $x_c$ are close to the known value $ 1/\mu.$ For $y \le 0.7$ this is clearly evident from the analysis. And as $y$ gets smaller still, so that  walks ending near to the surface are favoured, it is clear that the exponent is approaching $\gamma_{1,1}=-3/16=-0.1875$ as expected from the scaling law \cite{BGMTW78} $2\gamma_1-\gamma_{11}=\gamma+\nu.$  This is the exponent appropriate to arches, often called {\em loops} in the literature. They are walks in the half-plane whose origin and end-point both lie in the surface. For $y=0.8$ the series really does suggest that $x_c = 0.37918 \pm 0,00003$ with an exponent that looks very close to zero. For $y=0.9$ the approximants suggest that $x_c = 0.3792 \pm 0.0002,$ (so could be $1/\mu$) but there is another singularity very close by (at around $0.389$), and it is known that in such situations the estimate of the location, and exponent, is less trustworthy. 

This sort of behaviour is typical of the situation in series analysis when one is in the vicinity of discontinuous change in the critical exponent. The only way a finite series -- that is to say, a polynomial approximation to an infinite series -- can mimic this discontinuity is by shifting the critical point slightly. So our conclusion is that the observed behaviour is consistent with the known result that $x_c = 1/\mu$ for $y \le 1,$ and that the exponent changes discontinuously from $-\gamma_1 = -0.953125$ to $-\gamma_{1,1} = 0.1875$ as $y$ decreases below 1. 

For $y >1.04$ the series are beautifully behaved, the singularity is clearly seen to be a simple pole, and we can provide 10 digit (or more) accuracy in estimates of the critical point.
For $1 < y < 1.04$ we get the sort of behaviour we expect with a discontinuous change in exponent as we transition from an exponent $\gamma_1 \ne 1$ to a simple pole.

So, in summary, it appears that for $y < 1$ we have $x_c = 1/\mu$ and exponent $\gamma_{1,1} = -3/16;$  for $y=1$ we have $x_c=1/\mu$ and exponent $\gamma_1 = 61/64$ and for $y > 1$ we have $x_c$ monotonically decreasing as $y$ increases, and with a simple pole singularity.

An interesting and unexpected feature is that for all values of $y$, the location of the {\em antiferromagnetic} singularity -- that is to say, the singularity on the negative real axis, which for unconstrained SAWs is at $x=-1/\mu,$ --  is unchanged at $-1/\mu$ with exponent $3/2$\footnote{To be precise, this singularity is less obvious for $y \ge 4.$ This can be understood from the fact that the radius of convergence of the series decreases as $y$ increases. The location of the anti-ferromagnetic singularity is located at a distance more than twice the radius of convergence when $y \ge 4$, so is increasingly difficult to detect. One would have to expect that it is there nonetheless.}.

A further bonus of this analysis is that the series analysis is exquisitely sensitive to the value of $y$ near $y=1.$ This gives us a method for confirming that $y_c=1$ \cite{IoffeVelenik, Rensburg2009}. From table \ref{tab:a1}, giving the results of an analysis of the series  $H(x,1)$ (see the second column of table~\ref{tab:y1}) we find a value of the critical point very close to $1/\mu.$ The value of $1/\mu$ is 0.379052277751, with uncertainty in the last digit only \cite{Clisby12}. We can vary our estimate of $y_c$ until we get agreement with $1/\mu,$ and this turns out to be at $y_c=0.9999995 \pm 0.0000005.$  We know that
$ y_c \ge 1$, so combining our numerical results with this rigorous result, we conclude that $y_c=1$. So it seems that for $y=y_c$ the exponent is given by $\gamma_1,$ and that this changes discontinuously to a simple pole for $y > y_c.$ For $y < y_c$ the evidence strongly suggests that the exponent is given by $\gamma_{1,1}.$

\begin{table}
   \centering
   \begin{tabular}{|l|l|l|}
   \hline
      $y$    & $x_c$ & Exponent\\
\hline
   0.4 & 0.379053 & 0.186 \\
0.5 & 0.379052 & 0.186 \\
0.6 & 0.379052 & 0.187 \\
0.7 & 0.37905 & 0.195 \\
   0.8      & 0.37918 & 0.00 \\
   0.9             & 0.3792   &  -0.3 \\
0.99 & 0.37925 & -0.63 \\
0.999 & 0.3790837 & -0.9328 \\
0.9999 & 0.379055 & -0.950 \\
0.99999 & 0.379052628 & -0.95296\\
0.999999 & 0.37905229 & -0.95307 \\
   1.0       & 0.37905225  & -0.95308 \\
1.000001 & 0.37905221 & -0.95309 \\
1.00001 & 0.37905188 & -0.95321 \\
1.0001 & 0.3790488 & -0.9547 \\
1.001  & 0.379019 & -0.970 \\
1.01 & 0.37862 & -1.11 \\
    1.02       & 0.37804  & -1.137 \\
      1.04 &0.37649   &  -1.0 \\
       1.06 &0.37463   &  -0.99 \\
        1.08 &0.37265   &  -0.99\\
        1.1 &0.370564   &  -1 \\
         1.2 &0.3592886   &  -1 \\
          1.3 &0.3475682   &  -1 \\
           1.5 &0.3249328   &  -1 \\
            1.75 &0.2995547603   &  -1 \\
 2.0 &0.2775487   &  -1 \\
      2.5 & 0.2418862105 & -1 \\
3& 0.214449855 & -1 \\
4 & 0.1751070033 & -1 \\
    5 & 0.14820871438 & -1  \\
7 & 0.1136573165016 & -1 \\
10 & 0.084421281924 & -1 \\
20 & 0.045635244067 & -1 \\
40 & 0.02383593409377 & -1 \\
60 & 0.01613729712 & -1 \\
90 & 0.01087210691 & -1 \\
150 & 0.00657951322 & -1 \\
250 & 0.003968378456 & -1 \\
  \hline    
   \end{tabular}
  \caption{SAWs at a surface. Estimates of $x_c$ for $a=1$  and various $y$ values. 
  For $y > 1$ the singularity is a simple pole.
For all $y$ values, there is also an anti-ferromagnetic singularity at $-1/\mu$ with exponent $1.5$.}
   \label{tab:a1}
\end{table}

\subsection{No applied force. $y=1.$}\label{sec:kappa}

Define the generating function
\begin{equation}
K(x,a) = \sum_n C_n(a,1) x^n=\sum_n e^{\kappa(a) n + o(n)} x^n.
\end{equation}
$K(x,a)$ will be singular at $x=x_c(a) = \exp[-\kappa (a)]$
and, close to this singularity,  $K(x,a)$ should behave as 
\begin{equation}
K(x,a) \sim \frac{B}{[x_c(a)-x]^{\gamma (a)}}
\end{equation}
where $\gamma (a)$ is a critical exponent whose value depends on the value of $a$.

\begin{figure}
\centering
\includegraphics[scale =0.5] {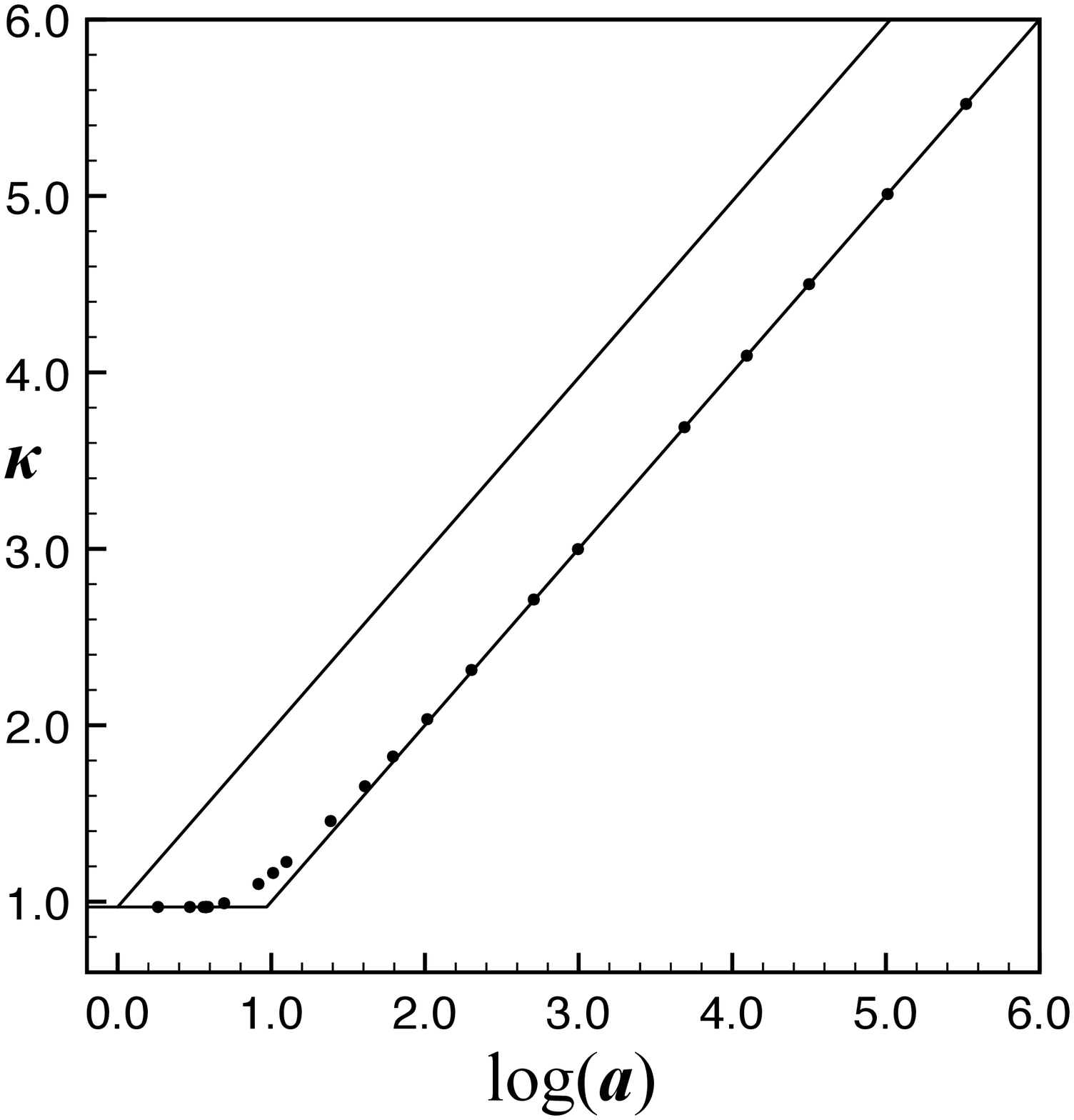}
 \caption{The $a$-dependence of the free energy $\kappa (a)$. The straight lines 
 indicate the exact lower and upper bounds, 
 where for $\log a < 0$ we know that $\kappa (a) = \log \mu$ while for $ \log a > 0$ we know that
$ \max[\log \mu, \log a] \le \kappa (a) \le \log \mu + \log a$.}
 \label{fig:kappa}
\end{figure}
We have analysed the series  $K(x,a),$ corresponding to the ``no force'' situation. As in the previous ``no interactions'' case,  we find that the series is exquisitely sensitive to the value of $a_c.$ The best estimate \cite{Beaton2012} currently is $a_c = 1.77564,$ with errors expected to be confined to the last quoted digit. That estimate is a comparatively recent result, which improved dramatically on pre-existing estimates, so improving on this, as we have done, is quite surprising. From  table \ref{tab:y1} below, we see from the second column that we find a value of the critical point very close to $1/\mu$. We can vary our estimate of $a_c$ until we get agreement with $1/\mu,$ and this turns out to be at $a_c=1.775615 \pm 0.000005$ with exponent $1.45395$ which is satisfyingly close to the conjectured exact value \cite{Guim1989, Batchelor1995} $\gamma_1^{sp}=93/64=1.453125,$ where the superscript refers to the ``special'' transition that takes place right at the adsorption temperature \cite{Guim1989}. As $a$ increases, we quickly see a simple pole emerging. So it seems that for $a=a_c$ the singularity is characterised by a (diverging) exponent $93/64,$ and that this changes discontinuously to a simple pole for $a > a_c.$ For $a < a_c$ the exponent is, as we would expect, given by $\gamma_1.$  
If we analyse the series with $y=0$ and $a=a_c,$ we can estimate the exponent $\gamma_{11}^{sp}.$ We did this and found $\gamma_{11}^{sp} = 0.816 \pm 0.006,$ in agreement with the expected value $13/16 = 0.8125$ \cite{DS86}.
In figure~\ref{fig:kappa}  we give our estimates of the free energy $\kappa (a) = -\log x_c$ as a function of $\log a$.

\begin{table}
   \centering
   \begin{tabular}{|l|l|l|l|l|}
   \hline
      $a$    & $x_c$ & Exponent& $x^*$ & Exponent\\
\hline
1.3 & 0.379052 & -0.952 & -0.37905 & 1.5 \\
1.6 & 0.379058 & -0.96 & -0.3793 & 1 \\
1.75 & 0.37910 &-1.31 & -0.37918 & 0.333 \\
   1.77559 & 0.37905237 & -1.4538 & -0.379050 & 0.256\\
1.775615 & 0.37905227 & -1.4539 & -0.379050 & 0.256 \\
  1.77564 & 0.37905217 & -1.4541 & -0.37905 & 0.256\\
  1.775665 & 0.37905207 & -1.4542 & -0.37905 & 0.256 \\
  1.77569 & 0.37905197 & -1.4538 & -0.37905 & 0.256\\
1.8 & 0.37893 & -1.58 & -0.378885 & 0.189 \\
2.0 & 0.37112 & -1.06 & -0.377 & -0.71\\
2.5 & 0.332682 & -1.0008 & -0.365065 & -0.517\\
2.75 & 0.3125387 & -0.999995 & -0.35806 & -0.5002 \\
3.0 & 0.293630848 & -1 & -0.35106 & -0.5005 \\
4.0 & 0.2329152160359 & -1 & -0.325298 & -0.499 \\
 5.0 & 0.191211527263626 & -1 & -0.30403 & -0.5005 \\
 6.0 & 0.16158981267578 & -1 & -0.28652 & -0.4995 \\
7.5 & 0.130751327296498 & -1 & -0.26538 & -0.4994 \\
 10.0 & 0.0989240104593583 & -1 & -0.23912 & -0.4996 \\
       15.0 & 0.0663536371608435 & -1 & -0.20474 & -0.496 \\
 20.0 & 0.049869446447162 & -1 & -0.18249 & -0.503 \\
40 & 0.0249840050794006 & -1 & -0.1367 & -0.6 \\
60 & 0.01666196255 & -1 & & \\
90 & 0.01110972448 & -1 & & \\
150 & 0.0066663684218 & -1 & & \\
250 & 0.00399993575 & -1 & & \\
        
  \hline    
   \end{tabular}
 \caption{SAWs at a surface. Estimates of $x_c$ for $y=1$  and various $a$ values. For 
 $a > a_c,$ the singularity is a simple pole.
There is also a second, antiferromagnetic singularity at $x=x^*$ with exponent $-1/2.$}
   \label{tab:y1}
\end{table}

The behaviour of the antiferromagnetic singularity is different from that observed in the previous sub-section. For $a < a_c$ it seems stable at $-1/\mu,$ with an exponent that is likely to be exactly $1.5,$ as for the case above. For $a > a_c$ however, the anti-ferromagnetic critical point monotonically decreases as $a$ increases for $a > a_c,$ and (conjecturally) has a square root singularity. At $a=a_c$ it looks more like a fourth root branch point, but a zero, not a divergence.\footnote{The estimate of the singularity location is not very precise, so it's entirely possible that the exponent is not exactly $1/4,$ but some fraction of approximately similar value.}

\subsection{Phase diagram calculation}

\begin{figure}
\centering
\includegraphics[scale =0.7] {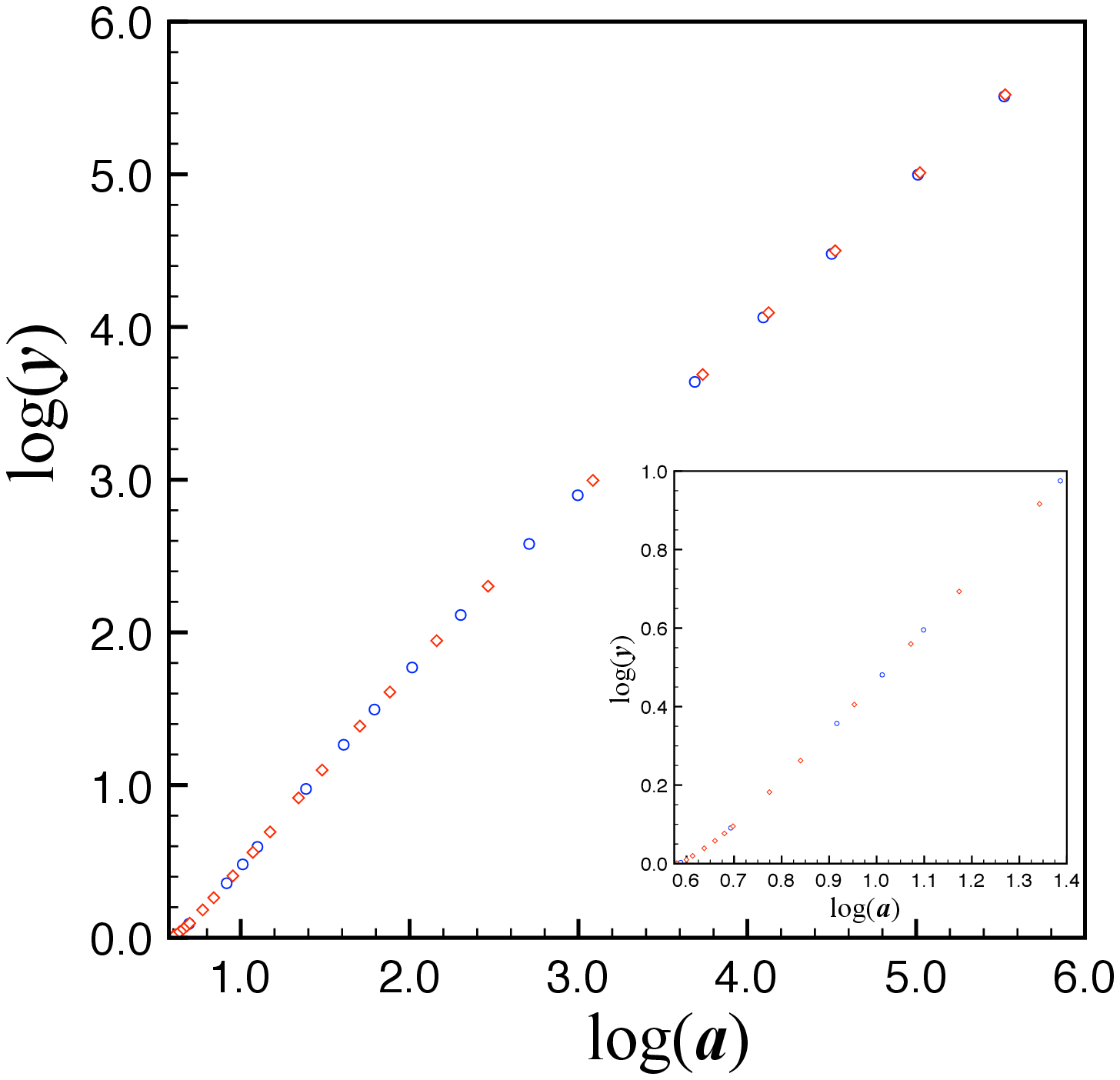}
 \caption{The phase boundary between the adsorbed and ballistic phases in the 
 $(\log{a},\log{y})$-plane. The blue circles correspond to the data from table~\ref{tab:ay1} while
 the red diamonds correspond to  the data from table~\ref{tab:ay2}. 
  The inset is a blow-up of the region near the origin.}
 \label{fig:phaseboundary}
\end{figure}

In order to locate the phase boundary between the adsorbed and ballistic 
phases, in the $(a,y)$-plane, we make use of (\ref{eqn:psi}).  $\psi(a,y)$ is equal to 
$\kappa (a)$ throughout the adsorbed phase and to $\lambda (y)$ throughout the 
ballistic phase.  The phase boundary between the adsorbed and ballistic phases is the 
locus of points where $\kappa (a) = \lambda (y)$.  For a given value of $a$ we calculated 
$\kappa (a)$ as in Section \ref{sec:kappa} and then found the value of $y$ such that 
$\lambda (y) = \kappa (a)$ by interpolating the results for $\lambda (y)$ found in 
Section \ref{sec:lambda}.  More precisely, from table~\ref{tab:a1}  we calculated $y=f_1(x_c)$ by using the program Eureqa \cite{Eureqa} on columns 1 and 2 of table~\ref{tab:a1}. 
As a relevant technical detail, in actual implementation it is desirable to minimize the variation of the parameters as much as possible. 
Accordingly, we sought a fit to the functional form $1/y={\tilde f}(x_c(1)-x)$ where $x_c(1)=1/\mu=0.3790522777\ldots.$
In this way we found an interpolation formula, which we used by
 inserting the $x_c$ values from table~\ref{tab:y1}, so as to obtain the $y$ values corresponding to the $a$ 
 values in table~\ref{tab:y1}. In this way we obtained the results shown in table~\ref{tab:ay1}.

As a check, we also calculated points on the curve starting with a 
given value of $y$ and reversing the procedure.  More precisely, from table~\ref{tab:y1} we calculated $1/a=f_2(x_c)$ also by using Eureqa on columns 1 and 2 of table~\ref{tab:y1}, and found an interpolation formula. We obtained a further set of $(a,y)$ values by substituting the $x_c$ values  from table~\ref{tab:a1}, so as to obtain the $a$ values corresponding to the $y$ values in table~\ref{tab:a1}. In this way we obtained the results shown in table~\ref{tab:ay2}. 

Combining the data in these two tables results in the phase boundary shown in 
figure~\ref{fig:phaseboundary} where we have plotted the data from table~\ref{tab:ay1} as red circles and the data 
from table~\ref{tab:ay2} as blue diamonds. The close agreement between the two independent analyses 
implies that the results are accurate to at least graphical accuracy..  The curve passes through the point $(a_c^o,y_c^o)$, is 
strictly monotone increasing and asymptotic to $y=a$, as shown in \cite{JvRW2013}.  It is 
interesting to note that the curve is not concave.

\begin{figure}[htbp]
   \centering 
 \includegraphics[scale=0.5]{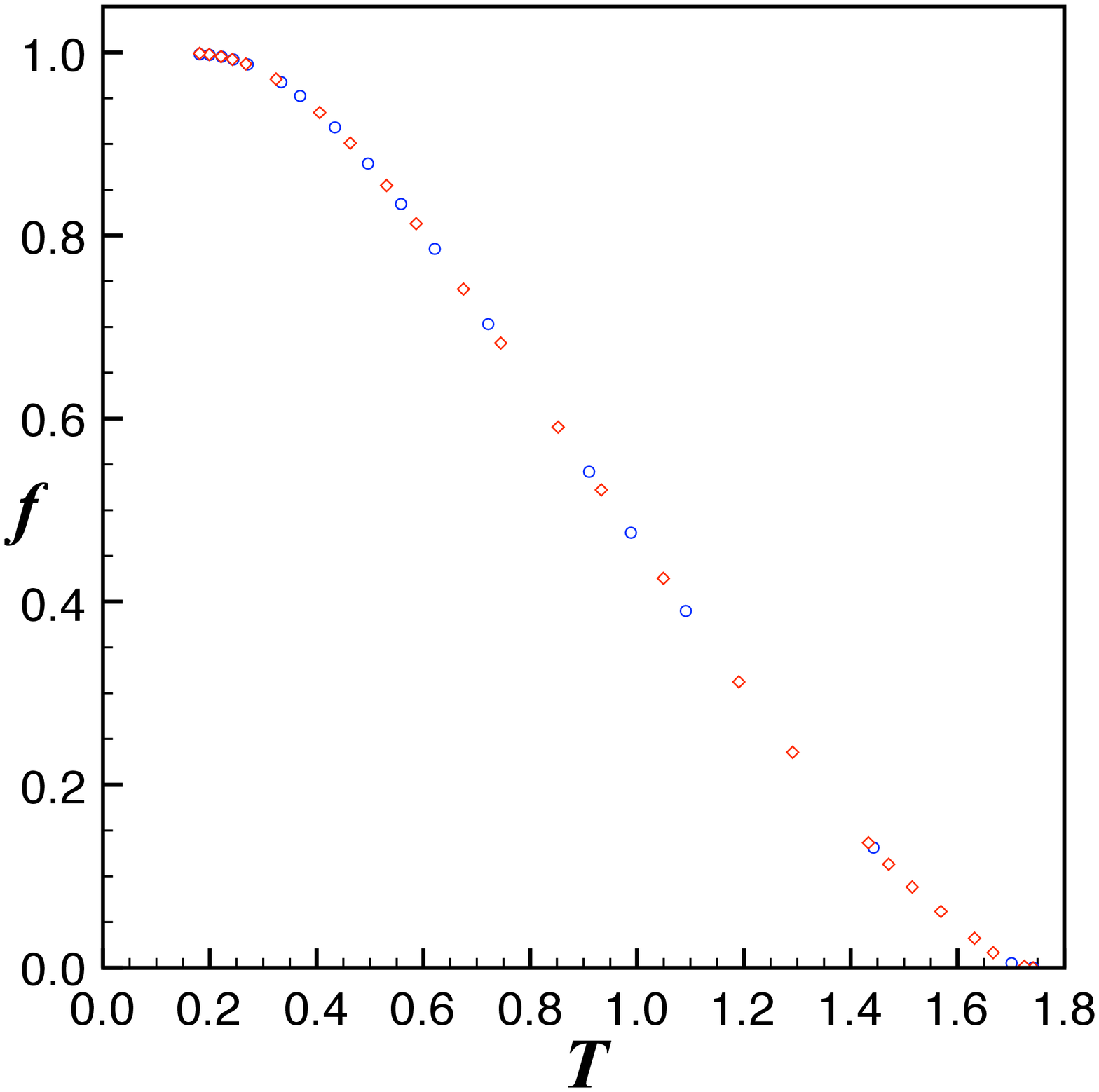} 
\caption{The phase boundary given as a force-temperature diagram. The horizontal axis is $T=\frac{1}{\log(a)}$, the vertical axis is the force, $f=\frac{\log(y)}{\log(a)}.$}
   \label{fig:ft1}
\end{figure}

We can switch to physical variables (force and temperature)
using (\ref{eqn:physvar}).  Without much loss of generality 
we can set $\epsilon = -1$ and work in units where $k_B=1$.  The corresponding phase
boundary in the force-temperature plane is given in figure~\ref{fig:ft1}.  Notice that
the force at zero $T$ is 1 and the limiting slope at $T=0$ is zero, as predicted in
\cite{JvRW2013}.  The curve is monotone decreasing as $T$ increases, with no re-entrance.
See for instance \cite{JvRW2013,Mishra2005,Skvortsov2009} for further discussion.
The force-temperature curve  is in semi-quantitative agreement with an earlier numerical study by Mishra \emph{et al}
\cite{Mishra2005}, but is substantially more precise.

\begin{table}
   \centering
   \begin{tabular}{|l|l|l|}
   \hline
      $a$    & $x_c$ &  $y(x_c)$\\
\hline
1.775615 & 0.37905227 &   1 \\
  1.77569 & 0.37905197 &1.0001\\
1.8 & 0.37893 &  1.0030 \\
2.0 & 0.37112 &  1.0953\\
2.5 & 0.332682 &  1.4293\\
2.75 & 0.3125387 &  1.6174 \\
3.0 & 0.293630848  &1.8137\\
4.0 & 0.23291521603  &2.6510 \\
 5.0 & 0.19121152726  &3.5399 \\
 6.0 & 0.16158981268 & 4.4592 \\
7.5 & 0.13075132730 & 5.8729 \\
 10.0 & 0.09892401046  & 8.2816 \\
       15.0 & 0.06635363716  & 13.189 \\
 20.0 & 0.04986944645 &  18.146 \\
40 & 0.02498400508 &  38.118 \\
60 & 0.01666196255 & 58.136 \\
90 & 0.01110972448  & 88.145\\
150 & 0.00666636842 & 148.00 \\
250 &   0.00399993575     &  247.18 \\
  \hline    
   \end{tabular}
 \caption{ Phase diagram estimated by calculating $y(x_c)$ from the interpolation formula found by Eureqa.}
   \label{tab:ay1}
\end{table}

\begin{table}
   \centering
   \begin{tabular}{|l|l|l|}
   \hline
      $y$    & $x_c$ & $a(x_c)$\\
\hline

   1.0       & 0.37905225  & 1.775615 \\
1.001  & 0.379019 &1.7854 \\
1.01 & 0.37862 & 1.8220 \\
    1.02       & 0.37804  & 1.8456 \\
      1.04 &0.37649   &  1.8915 \\
       1.06 &0.37463   &  1.9347 \\
        1.08 &0.37265   &  1.9733\\
        1.1 &0.370564   &  2.0092\\
         1.2 &0.3592886   &  2.1693 \\
          1.3 &0.3475682   &  2.3161 \\
           1.5 &0.3249328   & 2.5934 \\
            1.75 &0.2995547603   &  2.9206 \\
 2.0 &0.277548710  &  3.2329 \\
      2.5 & 0.2418862105 &3.8287 \\
3& 0.214449855 & 4.4003 \\
4 & 0.1751070033 & 5.5029 \\
    5 & 0.14820871438 & 6.5747  \\
7 & 0.11365731650 & 8.6708 \\
10 & 0.08442128192 & 11.757 \\
20 & 0.04563524407 & 21.876 \\
40 & 0.02383593409 & 41.923 \\
60 & 0.01666196255 & 61.912\\
90 & 0.01110972448 & 91.864 \\
150 & 0.00666636842 &151.74 \\
250 & 0.00399993575 & 251.50 \\

  \hline    
    \end{tabular}
  \caption{Phase diagram estimated by calculating $a(x_c)$ from the interpolation formula found by Eureqa.}
   \label{tab:ay2}
\end{table}

\section{The behaviour and nature of the phase transition on the phase boundary.}

It is possible to use the results of \cite{JvRW2013} to prove that the phase transition
from the ballistic to the adsorbed phase is first order.  We state this as a theorem.
\begin{theo}
The free energy $\psi(a,y)$ is not differentiable at the phase boundary between
the ballistic and adsorbed phases, except perhaps at the point $(a_c^o,y_c^o)$.
\end{theo}
\Pr
There is a 
monotone strictly increasing curve 
$y=y_c(a)$ in the $(a,y)$-plane, through the point 
$(a_c^o,y_c^o)$, corresponding to the phase boundary between the 
ballistic and adsorbed phases.  In the ballistic phase $\psi(a,y)=\lambda(y)$
and in the adsorbed phase $\psi(a,y)=\kappa(a)$.  The free energy $\kappa (a)$
is a monotone increasing function of $a$, convex in $\log a$.  It therefore has 
left and right derivatives at every value of $a$. 
Throughout the adsorbed phase the left and right derivatives of $\kappa (a)$
are positive.  Consider a line of fixed $y=y_1 > y_c^0 \ge 1$.  The free energy $\psi(a,y_1)=\lambda(y_1)$
for $a \le a_c(y_1)$ and $\psi(a,y_1)=\kappa(a)$ for $a \ge a_c(y_1)$. For $a \le a_c(y_1),$
$\partial \psi(a,y_1) /\partial a = 0$.  For $a \ge a_c(y_1)$ the right derivative of 
$\kappa (a)$ with respect to $a$ is positive.  Therefore the left and right derivatives
of $\psi(a,y)$ with respect to $a$ at $(a_c(y_1),y_1)$ are not equal and the free energy is not differentiable.
\qed
\vspace{4mm}
However it is still of interest to determine how the free energy behaves as we approach the phase boundary. We ``know" the location of the phase boundary from the results reported in the previous section, at least to graphical accuracy. That is to say, with an accuracy of three to four significant digits. At the phase boundary, we also know the value of the radius of convergence, from the data above. So, by way of example, at $a=2$ the phase boundary is at $y=1.095.$ If we analyse the series at this point, (recall this just means substituting the required values of  $a$ and $y$ into the three-variable generating function we have, which produces a one-variable generating function, where the expansion variable is conjugate to the length of the walk) we find the critical point is at $x_c = 0.371125$ with an exponent of $-1.995.$ This is exactly the same value of $x_c$ found at $a=2, \,\, y=1,$ though at $(2,1)$ the exponent is a simple pole. 

As we increase $y,$ we see exactly the same behaviour as observed in section 1 that allowed us to identify $a_c.$ That is to say, there is a variation in the criticial point and critical exponent as the approximants struggle to cope with a discontinuous exponent change. So at $(2,1.05)$ the $(x_c,exponent)$ pair is estimated to be $(0.37135, -1.36 \pm 0.3).$ This large error in the exponent estimate is a signature that the analysis method is struggling. At $(2,1.09)$ we find $(0.371325, -1.962 \pm 0.016),$ and at $(2,1.095),$ which is our best estimate of the intersection of the line $a=2$ with the phase boundary, we find for the critical point and exponent $(0.371125,-1.998 \pm 0.013).$ Note that this value of the critical point is exactly that found when $a=2$ and $y=1,$  which is well below the phase boundary.

Suggestive as this is, we note that the value of $a$ chosen is rather close to the point $(a_c,y_c)$
(a bicritical point, see \cite{Klushin1997}),  where the behaviour is different, and may still have an effect on the convergence rate of the series. So we take another example, when $a=3,$ which is well away from the point $(a_c,y_c)$. At $(3,1)$ the critical point is $x_c=0.2936308\ldots,$ and the exponent is a simple pole. The phase boundary point at $a=3$ is estimated to be at $y=1.8137.$ Analysis of the series at that point gives $(0.293636, -2.0001 \pm 0.0002),$ rather confirming the double pole. The location is slightly different from that observed at $(a=3, y=1),$ but that is likely ascribable to the estimate of $y$ on the phase boundary being slightly in error. Indeed, if we repeat the analysis with $y=1.8139,$ we find $x_c=0.293627\ldots,$ and the exponent estimate is $-2.0000018 \pm 0.0000022,$ which is rather convincing evidence for a double pole!

So we find that at (or very near to and below) the phase boundary, the series are well-converged, give the estimates of the critical point we expect, and display a double pole singularity. Near to the phase boundary, the estimates are very variable, and behave in exactly the same way as did the series analysed in section \ref{sec:results},  with the approximants seemingly struggling to cope with a discontinuous change in the critical exponent.

It is instructive to consider the simpler case of directed positive walks.  We discuss two cases:
\begin{enumerate}
\item
Positive walks with step set $(1,1)$ and $(1,-1)$, which we refer to loosely as Dyck
paths, and 
\item
Positive walks with step set $(1,1)$, $(1,-1)$ and $(1,0)$, which we refer to loosely as Motzkin paths.
\end{enumerate}

It is straightforward to solve each of these models exactly, though we don't give 
the details here.  They each show three phases (a free 
phase, an adsorbed phase and a ballistic phase).  There is a phase boundary between the 
adsorbed and ballistic phases and in both cases these phase boundaries are concave in
the $(\log a, \log y)$-plane.  In the Motzkin path case the phase boundary is asymptotic to $y=a$ while 
in the Dyck path case it is asymptotic to $y = a^{1/2}$, because a maximum of half the vertices 
can be in the surface.  In each case the singularity in both the adsorbed and ballistic phases is a 
simple pole so the generating function has both of these singularities though one is dominant
in the adsorbed phase and the other is dominant in the ballistic phase.  On the phase 
boundary between the adsorbed and ballistic phases these two singularities are equal resulting in a 
double pole, just as observed numerically for the case of SAWs.

\section{Discussion}
\label{sec:discuss} \setcounter{equation}{0}

We have considered a self-avoiding walk model of  polymer adsorption at an 
impenetrable surface where
\begin{enumerate}
\item
the walk is terminally attached to the surface,
\item
the walk interacts with the surface with an attractive potential, and
\item
the walk is subject to a force applied normal to the surface at the last vertex 
of the walk.
\end{enumerate}
For the square lattice we have used series analysis techniques to investigate 
the phases and phase boundaries for the system.  There are three phases, a free
phase where the walk is desorbed but not ballistic, an adsorbed phase where 
the walk is adsorbed at the surface and a ballistic phase where the walk
is desorbed but ballistic.  We have located the phase boundaries and proved that
the phase transition from the adsorbed to the ballistic phase is first order.  In addition
we have very precise values for the critical points for adsorption without a force and 
for the free to ballistic transition with no surface interaction.

\section*{Acknowledgements}
The authors would like to acknowledge helpful conversations with Enzo Orlandini,
and help with the preparation of the data files by Jason Whyte.
This research was partially supported by NSERC of Canada.  
The computations for this work were supported by an award to IJ under the
Merit Allocation Scheme on the NCI National Facility  at the 
Australian National University.  
IJ and AJG were supported under the Australian Research Council's Discovery Projects 
funding scheme by the grants  DP120101593 and DP120100939 respectively.
We thank Cornell Creative Machines Lab for making available the program Eurequa \cite{Eureqa}.

\end{document}